# Phonon-induced topological insulating phases in group IV-VI semiconductors

*Jinwoong Kim and Seung-Hoon Jhi* [*]

Department of Physics, Pohang University of Science and Technology, Pohang 790-784, Republic of Korea

## ABSTRACT

Development of topological insulating phases in IV-VI compounds under dynamic lattice deformations is studied using first-principles methods. Unlike the static state of topological phases at equilibrium conditions, we show that non-trivial topological phases are induced in the compounds by the dynamic lattice deformations from selective phonon modes. Calculations of the time-reversal polarization show that the $Z_2$ invariant of the compounds is flipped by the atomic displacements of selective phonon modes and that the compounds exhibit oscillating topological phases upon dynamic lattice deformations. Our results indicate that the elementary excitations in solids can trigger topological phases in trivial band insulators.

Discovery of three-dimensional topological insulator (TI) casts a new frame to classify solids.[1-6] When the time-reversal symmetry is preserved, insulators are classified as either topological insulators or trivial insulators according to their topological invariant $Z_2$ number.[1-3] Since the topological invariant cannot be changed without going through the energy-gap closing, topological insulators should develop conducting helical states at the surfaces or the interface to trivial insulators. Because of the spin-momentum locking of the helical states, back-scattering-free transport is feasible along the surfaces or interfaces. In this respect, the TI surfaces can work as excellent conducting channels of spin-polarized currents, reserving great potentials in spintronics applications. In addition, TIs are also expected to serve as low-energy platforms to detect exotic particles proposed in high energy physics. For example, the Majorana fermion, which refers to particles that are their own anti-particles, was suggested to exist at the interface of topological insulator and superconductors.[7, 8] It will be more efficient to prepare hetero-phase domains in a single material; for example, TI phase in a superconductor or vice versa in realizing such exotic particles.[9, 10] Manipulating topological phases by controlling materials parameters thus opens to explore intriguing phenomena as well as to develop TI-based electronic devices.

Many emerging phenomena in solids are explained by the notion of elementary excitations and interactions between them. While topological phases are usually referred to as geometrical characteristics of electronic structures of solids at equilibrium conditions, it is also of great physical interest to explore whether elementary excitations or interactions between them can lead to the development of topological phases or to new emerging phenomena associated with topological phases. PbTe, SnTe, GeTe, and PbS, are narrow-gap semiconductors and known to exhibit superconducting phases upon pressure[11] or doping[12, 13]. Recently, the topological-crystalline-insulating phase is also found in this group of compounds.[14-16] It is very viable in these compounds that interplay of crystalline symmetries and the time-reversal symmetry leads to very rich behavior of topological phases. In this Letter, we studied emerging topological phases in PbTe, SnTe, GeTe, and PbS under dynamic lattice deformations by selective phonon modes using first-principles calculations. This study provides an interesting demonstration that elementary excitations can induce non-trivial topological phases in narrow-gap semiconductors.

First-principles calculations were carried out using Vienna ab-initio simulation package (VASP).[17] The exchange-correlation of electrons is treated within the generalized gradient approximation (GGA) as parametrized by Perdew, Burke, and Ernzerhof (PBE).[18] Since band gaps calculated with GGA typically underestimate experimental values, we also used the screened hybrid-functional (HSE06)[19] to correct the band gaps (see Supplementary Materials). The projector-augmented-wave (PAW) method[20, 21] is used for atomic pseudopotentials and the plane-wave basis are expanded up to a cutoff energy of 300 eV. The Brillouin zone is sampled with 8×8×8 k-point grid for the bulk and 4×4×1 k-point grid for the slab structures. In both cases, Γ-point is included in the self-consistent calculations in order to properly take into account the effect of the band inversion at high-symmetry k-points. The spin-orbit interaction is included as implemented in the VASP code. We chose the parity-check scheme proposed by Fu and Kane for the verification of the $Z_2$ class.[22] For surface-band calculations, anion (Te or S)-terminated slab structures are used with the structural parameters obtained in the bulk calculations. The slab is constructed to have 73 monoatomic layers along the [111]-direction and the dangling bonds at surfaces are passivated by hydrogen atoms.

First, we studied the effect of lattice deformations by hydrostatic pressures and uniaxial strains. The time-reversal polarization (and the $Z_2$ invariant) is calculated from the parity (or the band inversion) at the time-reversal invariant momenta (TRIM).[22] We particularly focused on the $L$ points, where the band inversion occurs upon external pressure.

There are four equivalent $L$ points in the Brillouin zone, and we considered uniaxial strains along [111] direction to remove the degeneracy of the $L$ points. Upon the strains, the rock-salt structure of the compounds changes to the rhombohedral structure and one $L$ point in the [111] direction turns into $T$ point that is distinguished from other three $L$ points [Fig.1 (a) and (b)]. Figure 1(c) illustrates the behavior of the band-gap and the band inversion at $L$ points for PbTe under hydrostatic pressures (also see Supplementary Materials). When the band inversion occurs, the parity state at TRIMs ($\Gamma$, $L$, and $X$ points) changes from $\{1\Gamma^-, 4L^-, 3X^-\}$ to $\{1\Gamma^-, 4L^+, 3X^-\}$. Here the number and the superscript in the TRIMs represent the number of equivalent k-points and the parity-product (half of occupied states), respectively. The parity-product at $L$ points changes but the $Z_2$ number does not change because of the degeneracy of the $L$ points. PbTe thus still remains as trivial insulator.[3, 22] However, non-trivial topological phases can be induced when the uniaxial strains along [111] directions are applied. Again the band inversion occurs at three $L$ points above a certain uniaxial strain (about 4.5% or $c/c_{eq}$=0.955 for PbTe) but the band-gap at $T$ point (one of four $L$ points is now distinguished becoming $T$ point) hardly changes under the uniaxial strains [Fig. 1(d)]. The parity state is now $\{1\Gamma^-, 3L^+, 1T^-, 3X^-\}$ and then PbTe comes to have a non-trivial topological phase.

The hydrostatic pressure and the uniaxial strains can be combined to control the topological phases. Calculated phase diagrams of the parity state ($\delta_i$) and corresponding topological phases for PbTe, SnTe, GeTe, and PbS are shown in Fig. 2 upon variation of the lattice constants, $a$ and $c$, together with deformation paths. For example, the uniaxial compressive strain along [111] direction deforms PbTe through the path $p_c$ and PbT experiences the band inversion only at three $L$ points but not at $T$ point to have a non-trivial topological phase (for surface band structures, see Supplementary Materials). Other compounds exhibit similar behavior in topological classes upon applied strains except for SnTe that has the band inversion at $L$ point even at equilibrium. We note that two different topological classes[3, 22] arise in GeTe and PbS upon the strains; the parity state of {$1\Gamma^-$, $3L^+$, $1T^-$, $3X^-$} corresponding to 1;(000) class and the other is the parity state of {$1\Gamma^-$, $3L^-$, $1T^+$, $3X^-$} corresponding to 1;(111) class. Particularly interesting deformation is for PbTe along the path denoted by $p_a$ [radial compressive strain in the (111) plane] that may open a possibility of hosting Majorana fermions. Because the indirect-gap is zero for these strains, PbTe is actually topological semimetal and its surface states overlap with bulk conduction bands. We note that this has very significant implication for realizing Majorana fermions. It is known that PbTe becomes superconducting phase above the critical pressure where the superconducting gap is stimulated by the hole pockets [13] and band inversions [23]. Superconducting carriers in strained PbTe thus have pairing-gap in the bulk, and they can simultaneously exist in the helical edge states. These gapless edge modes in superconducting PbTe are the Majorana fermionic bound states.[7] The hybrid structures of superconductor-topological insulator using PbTe with proper strains will serve as platforms to detect the exotic particles.

Now we consider dynamic lattice deformations by phonon modes that lower the structural symmetry. The atomic displacements of particular phonon modes that cause the parity-reversal the same way as the asymmetric strains is of our particular interest. If the parity-reversal leads to a change in $Z_2$ number, we should observe non-trivial topological phases and thus helical edge states for a short period of time. This is a reasonable expectation within the notion of the Born-Oppenheimer approximation. We chose the ionic displacements of the phonon modes at $L$ point, which double the unit-cell along the [111] direction. We calculated the band-gap at $\Gamma$ and $X$ points (which are projection of four $L$ points of the original single unit-cell) and the band inversion as a function of the ionic displacement. Figure 3 shows our calculated results for PbTe and PbS. For longitudinal acoustic (LA) modes, the band-gaps at $\Gamma$ and $X$ points decrease as the atomic displacements are increased. In a small window of atomic displacements of around 0.2 Å, the parity state becomes {$1\Gamma^-$, $3L^-$, $1T^-$, $3X^+$} and the non-trivial topological phase appears. For longitudinal optical (LO) modes, the band-gap exhibits opposite movements at $\Gamma$ and $X$ points upon increasing ionic displacements. When the band inversion occurs only at three $X$ points above the displacement of about 0.3 Å for PbTe (0.15 Å for PbS), we observe a non-trivial topological phase emerging. We note that the fundamental band-gap still remains finite and these compounds are thus topological insulator in contrast to the case of LA modes.

In the case of the transverse modes, two independent (orthogonal) displacements should be considered because the modes are doubly degenerate. Upon ionic displacements, three equivalent $X$ points are now distinguishable as denoted by $X_i$ ($i=1,2,3$). Non-trivial topological phases appear if the parity at one of the $X$ points differs from those at other two $X$ points for transverse acoustic (TA) modes. When the ionic displacements along [$11\bar{2}$] direction are large enough with their energy exceeding 68 meV per formula unit, PbTe has the parity state of {$1\Gamma^-$, $3L^-$, $1T^-$, $X_1^+$, $X_2^-$, $X_3^-$} and thus a non-trivial topological phase. Because the band-gap reopens after the band inversion, PbTe is actually topological insulator [Fig. 3(c)]. Another parity state of {$1\Gamma^+$, $3L^-$, $1T^-$, $X_1^+$, $X_2^+$, $X_3^-$} is also possible for non-trivial topological phase if the band inversion at $\Gamma$ point occurs. In between the ionic displacements for the two non-trivial phases, the band inversion occurs only at two of the $X$ points and PbTe becomes weak topological insulator having two Dirac cones.

PbS (also other IV-VI compounds) exhibits similar behavior of emerging topological phases by phonon modes. We note that PbS has advantages over other compounds in experimental implementation of phonon excitations. The energies of ionic displacements to induce the topological phases are calculated to be about 58 meV per formula unit in TA mode, smaller than those for PbTe. In addition, the frequencies of phonon modes at $L$ point are well separated from each other; LA (3.16 THz), LO (6.90 THz), TA (1.53 THz), and TO modes (5.65 THz). This separation of the phonon frequencies is expected to be superior in preventing the interference of phonon modes. Once specific phonon modes are excited by the laser or using the impedance matching technique in phonon transmission from vibrating substrates, the topological phases can be detected by the pump-probe methods.

In order to confirm explicitly the topological phase upon ionic displacements, we calculated surface band structures of PbS with ionic displacements of the TA phonon mode (Fig. 4). The $X_i$ point in the bulk Brillouin zone is projected into $\bar{M}_i$ point of the surface Brillouin zone. For the atomic displacement along $[1\bar{1}0]$ direction [the path $d_1$ in Fig. 3(g)], the band-gaps at $\bar{M}_2$ and $\bar{M}_3$ points decrease as the atomic displacements are increased as shown in Fig 4(b). The Dirac cones are formed at $M$ points when the displacement is larger than the threshold value of ~0.5 Å (larger than for the bulk case due to the quantum confinement effect in the band gap). PbS becomes weak topological insulator [topological class of 0;(110)] for this atomic displacement along $d_1$. On the other hand, for a circularly-polarized TA phonon mode with ionic displacements along $d_2$ in Fig. 3(g), the band-inversion occurs sequentially at $\bar{M}_3$, $\bar{M}_2$, and $\bar{M}_1$ points. As a consequence, the Dirac cons form at the band-inversion $M$ points in the same series as shown in Fig. 4 (d) and (e).

Such a flickering formation of Dirac cones by lattice vibrations leads to intriguing physical consequences. For example, electrons that migrate back and forth to the surface to fill the Dirac-cone states produce electron-induced forces, which can be utilized to control atomic migrations (see also Supplementary Materials). This phenomenon will also increase the coupling between the surface-localized states and the ones in the bulk because the bulk should absorb the surface states at every extinguishing moment when the topological phase disappears due to the band-inversion recovery. This process can be used as a pump that transfers some spatial properties such as magnetic moment from one surface to the other. Direct measurement of the oscillating behavior may be possible by the time-resolved spectroscopy. A major concern to realize dynamic topological phases is large phonon energy required to generate the band inversion. One solution is to control the band-gap by well-known substitution methods[16] to reduce the required energy. Inducing topological phases (or turning it off) by specific phonon modes is also possible in other materials like tetradymite compounds ($Bi_2Te_3$, $Bi_2Se_3$, $Sb_2Te_3$, and even $Sb_2Se_3$). For tetradymite compounds, the band inversion is sensitive to the distance between quintuple layers that is associated with a phonon mode in which the thickness of quintuple layer vibrates. We expect that topological phase transition arises if particular phonon modes are excited with enough amplitude to cause the band inversion.

In summary, we studied non-trivial topological phases in group IV-VI semiconductors induced by lattice deformations. By lowering the crystal symmetry to remove the degeneracy of certain time-reversal invariant momenta, we showed that the time-reversal polarization can be controlled so as to induce the topological phases in the compounds. We discussed that these compounds can have multiphases of superconducting and topological-semimetallic states to serve as platforms to realize Marjorana fermions. By calculating the dynamic $Z_2$ invariant of the compounds by ionic displacements of selective phonon modes, we showed that dynamic topological phases and consequent flickering surface states can generate various transient phenomena on the surface of topological matters. Our study demonstrates that elementary excitations such as phonons can trigger the topological phases and generate intriguing transient phenomena in semiconductors.

Acknowledgement

This works was supported by the National Research Foundation of Korea (NRF) grant funded by the Korea government (MEST) (SRC program No. 2011-0030046). The authors would like to acknowledge the support from KISTI supercomputing center through the strategic support program for the supercomputing application research (No. KSC-2010-C2-0008).

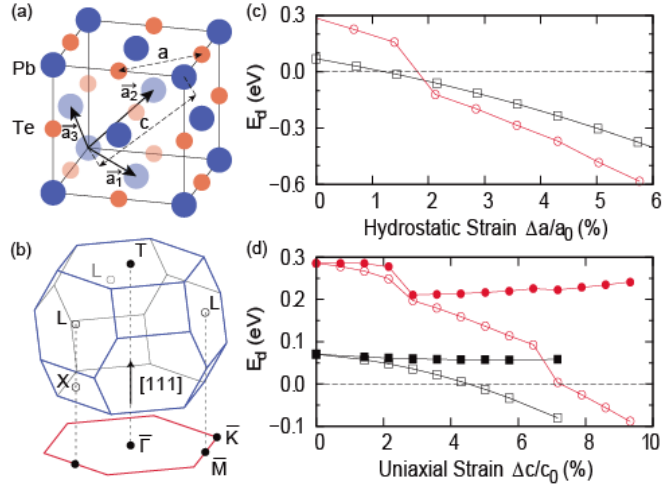

FIG. 1 (Color online) (a) Atomic structure of IV-VI compounds in the conventional unit cell. The arrows denote the primitive lattice vectors, and two lattice constants, *a* and *c*, are represented with dashed arrows. (b) The Brillouin zone of the rhombohedral structure and the time-reversal invariant momenta. The *T* point along the [111] direction is now distinguished from other three *L* points in the rhombohedral structure. The surface Brillouin zone projected into the (111) plane is shown below. (c) and (d), Calculated direct band-gaps of PbTe with GGA (square) and the hybrid functional HSE06 (circle) upon hydrostatic pressures and uniaxial stains, respectively. Filled symbols in (d) represent the data for *T* point as it is distinguished from *L* points under uniaxial strains. Negative band-gap means the band inversion. Here the uniaxial strains are applied under the constraint of $\left[\partial E/\partial a\right]_c = 0$ with *E* being the total energy.

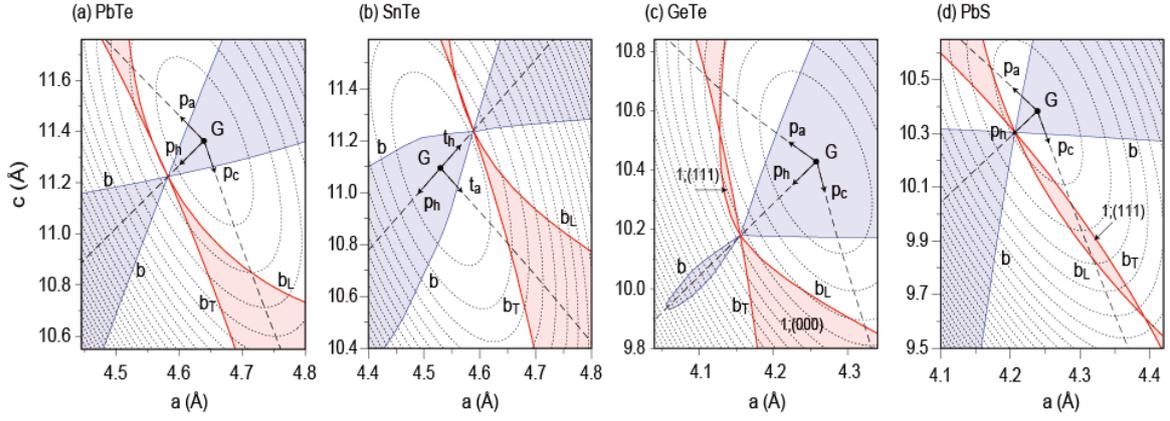

FIG. 2 (Color online) Calculated phase diagram of the $Z_2$ invariants for (a) PbTe, (b) SnTe, (c) GeTe, and (d) PbS upon lattice deformations [$a$ and $c$ as defined in Fig. 1(a)]. Dotted lines are the energy-contours with G representing the equilibrium lattice constants. The solid lines $b_L$ and $b_T$ denote the lattice parameters having the band inversion (or the parity change) at $L$ and $T$, respectively. The direct band-gap is finite in the whole range of $a$ and $c$, but the indirect-band gap is non-zero only in the blue region enclosed by the solid lines $b$. The compounds are thus topological semi-metal in the red region enclosed by $b_L$ and $b_T$ where the parity state is either $\{1\Gamma^-, 3L^+, 1T^-, 3X^-\}$ or $\{1\Gamma^-, 3L^-, 1T^+, 3X^-\}$. The deformation path for a particular strain is denoted by a dashed line with an arrow, each labeled as $X_Y$ ($X=\{p, t\}$; $Y=\{h, c, a\}$) where $X$ is for either compressive ($p$) or tensile ($t$) strain, and $Y$ is for the hydrostatic pressure ($h$), the uniaxial strain along the $c$-axis ($c$), or the radial strain normal to the $c$-axis ($a$) [for instance, $t_a$ corresponds to the stretching of the (111) plane].

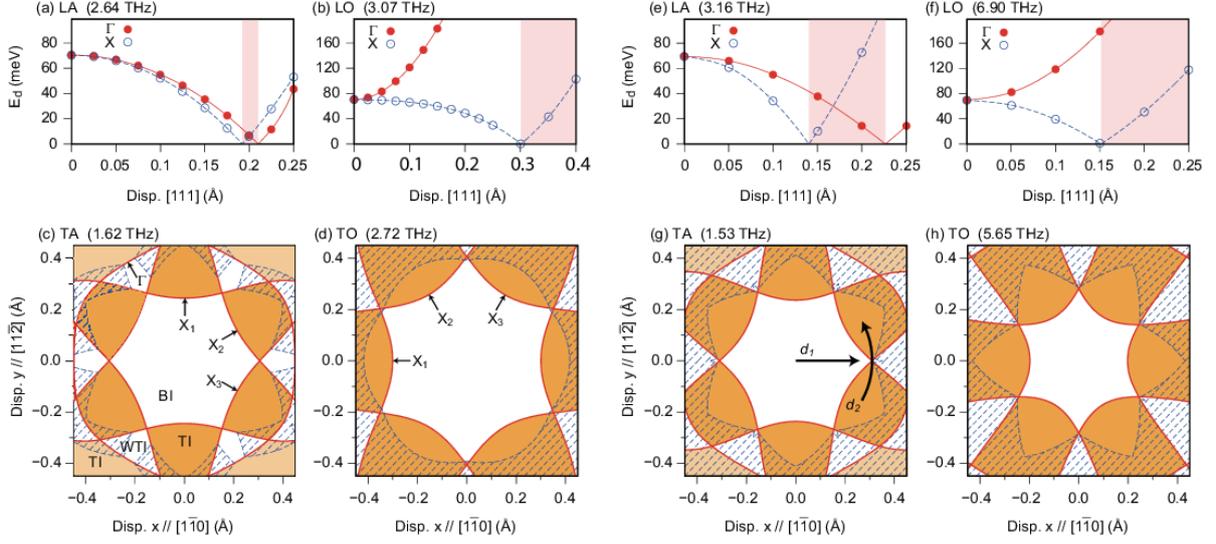

FIG. 3 (Color online) (a-d) Calculated band-gaps and the parity reversal upon atomic displacements of the phonon modes at *L* point for PbTe; (a) longitudinal acoustic (LA), (b) longitudinal optical (LO) modes, (c) transverse acoustic (TA), and (d) transverse optical (TO) modes. (e) – (f), the same plots for PbS. For longitudinal modes, the $Z_2$ invariant is changed when the band-gap at *X* point is closed but not at *Γ* point (the region in red). For transverse modes, the solid (red) lines denote the displacements of band-inversion at *Γ* or *X* points. When the ionic displacements cross the solid red lines by an odd (even) number of times, strong (weak) topological phase appears; inside the orange-colored (dark gray) region, it has one band inversion at $X_i$ point. Whereas, the band inversion occurs at two $X_i$ points and one *Γ* point (at three points in total) if the displacement is in the pale orange-colored (gray) region. For both cases, the compound becomes strong topological insulator and turns into weak topological insulator in other regions. Direct band-gap is finite in all regions except for the moment of the band inversion. However, the indirect band gap is zero inside the blue-dashed regions and the compounds become semimetallic.

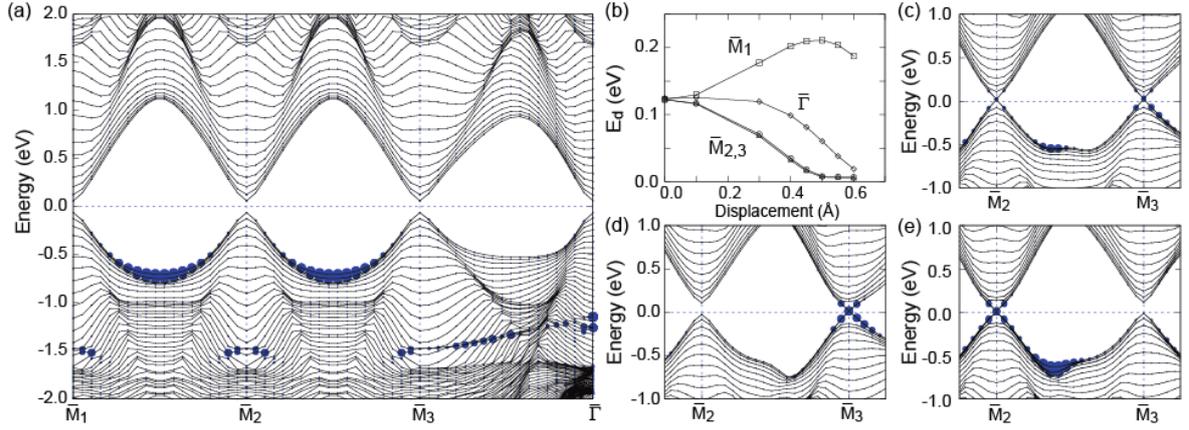

FIG. 4 (a) Surface band structure of PbS with no atomic displacement. (b) Calculated band gaps upon atomic displacements along the path $d_1$ denoted in Fig. 3(g). (c) Two Dirac cones are formed at $\bar{M}_2$ and $\bar{M}_3$ points for the displacement of 0.5 Å (PbS is then a weak topological insulator). On the other hand, the atomic displacement along the path $d_2$ in Fig. 3(g), which corresponds to the circularly-polarized TA phonon mode, allows only one Dirac cone at $\bar{M}_i$ point sequentially, as shown in (d) and (e), where the band inversion occurs. The Fermi level is set at the zero energy and the (blue) dots represent the states originating mostly from the surface atoms.